\documentclass[12pt]{article}
\usepackage[pdftex]{hyperref,graphicx,color}
\newcommand\farcs{\mbox{$.\!\!^{\prime\prime}$}}

\begin{document}
\noindent {\Large{\bf GBT Memo \#296}}
\begin{center}
\ \\
{\Large {\bf The GBT Beam Shape at 109 GHz}}
\ \\
David T. Frayer (Green Bank Observatory)\\
2017 March 31
\end{center}

\begin{abstract}

With the installation of the Argus 16-pixel receiver covering 75-115
GHz on the Green Bank Telescope (GBT), it is now possible to
characterize the antenna beam at very high frequencies, where the use
of the active surface and out-of-focus holography are critical to the
telescope's performance.  A recent measurement in good weather
conditions (low atmospheric opacity, low winds, and stable night-time
thermal conditions) at 109.4 GHz yielded a FWHM beam of $6\farcs7
\times 6\farcs4$ in azimuth and elevation, respectively.  This
corresponds to $(1.16 \pm 0.03)\,\lambda$/D at 109.4 GHz.  The derived
ratio agrees well with the low-frequency value of
$(1.18\pm0.03)\,\lambda$/D measured at 9.0 GHz.  There are no
detectable side-lobes at either frequency.  In good weather conditions
and after applying the standard antenna corrections (pointing, focus,
and the active surface corrections for gravity and thermal effects),
there is no measurable degradation of the beam of the GBT at its
highest operational frequencies.

\end{abstract}

\section{Introduction}

As of 2017, the GBT\footnote{The Green Bank Observatory is a facility
  of the National Science Foundation under cooperative agreement by
  Associated Universities, Inc.} has three available instruments
operating within the 3mm atmospheric window: the 4mm Receiver [1], the
Mustang-2 bolometer camera [2], and Argus -- the 16 element 75--115
GHz array [3].  The performance of the GBT within the 3mm band depends
significantly on winds and other environmental effects [4,5], tracking
errors, focus errors, errors associated with the active surface
corrections, and even systematic deformations of the panels [6].  The
weather and instrumental effects could be expected to degrade the
performance of the GBT with increasing frequency, since the errors
become a larger fraction of the wavelength and the expected beam size.
The purpose of this memo is to provide an estimate of the beam size
using Argus near 110 GHz, since the beam of the GBT has not been
previously characterized above 90 GHz.

\section{Beam Size Below 100 GHz}

For an idealized radio telescope assuming a Gaussian beam
approximation and adopting the standard GBT feed taper of 15dB (which
also corresponds to the feed taper of the Argus horns at 100 GHz [7]),
the expected beam size in radians is approximately $\theta_{\rm FWHM}
\approx 1.22\,\lambda/{\rm D}$ [8], where $\lambda$ is the observed
wavelength and D is the diameter of the GBT (100m).  To facilitate
comparisons at different wavelengths, measurements in this memo are
given in $\lambda/{\rm D}$.  Observers using Mustang/Mustang-2 have
reported typical beam sizes of $1.3\,\lambda/{\rm D}$ at 90 GHz~[2],
while the beam size at 77 GHz for the 4mm Receiver has been measured
as $1.34\,\lambda/{\rm D}$~[9].  This measurement was done before the
active surface model was updated in the fall of 2014~[10] and before
the new GBT servo-drive system was deployed in the summer of 2016.  In
summary, observations from previously seasons with the GBT within the
3mm band have been close to the theoretical expectation of $\approx
1.22\,\lambda/{\rm D}$, but were found to be typically slightly higher
($\sim 1.3\,\lambda/{\rm D}$).

For comparison with measurements at high frequency, I first measure
the beam at lower frequency in X-band (9.0\,GHz).  At this frequency,
the GBT active surface system is still used, but the instrumental and
weather effects on the size of the beam are expected to be minimized.
The weather was excellent for X-band at the time of these
observations; the data were taken during a scheduled Argus program to
provide the initial pointing solutions (AGBT17A\_423\_01, 2017.03.23,
PI A. Lovell).  Table~1 shows the beam size measured from the standard
azimuth and elevation peak scans used to derive the pointing offsets.
Averaging the azimuth and elevation results gives a beam size
measurement of $(1.18\pm0.03)\,\lambda/{\rm D}$ at 9.0\,GHz.  This is
slightly better than the theoretical expectation of
$1.22\,\lambda/{\rm D}$ for the Gaussian approximation.  Figure~1
shows the normalized beam profile in decibels (db $=10\log_{10}{\rm P})$ for the elevation scan.  This figure highlights
the excellent performance given by the unblocked optics of the GBT.
The peak of the beam profile is well approximated by a Gaussian, but
the wings of the beam profile falls off steeper than that of a
Gaussian and are better approximated by a Bessel function.  The first
side-lobe for the Gregorian focus position of the GBT is expected to
be -27dB lower than the peak~[10] and is not detectable in these data.
The measured X-band beam size of $1.18\,\lambda/{\rm D}$ is
significantly better than the typical values of 1.3$\,\lambda/{\rm D}$
previously observed above 66 GHz.

\section{Beam Size Above 100 GHz}

To characterize the beam properties above 100 GHz with the GBT, I used
data collected at 109.4\,GHz during Argus observations in good weather
conditions (AGBT17A\_304\_06\_01, 2017.03.23, PI A. Kepley).  These
data were collected on the same day as the comparison X-band
observations presented here.  The Argus data were taken after AutoOOF
observations with the 4mm Receiver to optimize the active surface for
the current thermal conditions.  The OOF solution returned an rms of
only $102\mu$m.  This is better than the typical OOF ``rms'' values of
of $\sim 200\mu$m, implying that only small corrections were needed
with respect to the default zernike model of the GBT [11] at the time
of observations.  Table~1 shows the beam size measured from the
azimuth and elevation peak scans. The source observed was the bright
radio quasar 3C273, which is estimated to be about 10\,Jy~[12].
Averaging the azimuth and elevation results gives a beam size
measurement of $(1.16\pm0.03)\,\lambda/{\rm D}$ at 109.4\,GHz.  The
result for Argus is consistent within uncertainties with the X-band
results.  Figure~2 shows the normalized beam profile in dB for the
elevation scan at 109.4\,GHz.  This is an impressive result for the
GBT system.

%
\begin{table}
\centering
  \caption{Beam Measurement Results}
  \begin{tabular}{lll}
    \hline \hline
 & X-band & Argus \\ 
 & 9.0 GHz& 109.4 GHz\\
\hline
FWHM (Az) & $80\pm2$ arcsec & $6.7\pm0.2$ arcsec\\
FWHM (El) & $82\pm2$ arcsec & $6.4\pm0.2$ arcsec\\
Avg (Az,EL)&$(1.18\pm0.03)\,\lambda/{\rm D}$
&$(1.16\pm0.03)\,\lambda/{\rm D}$\\
\hline
 \end{tabular}
\end{table}

\section{Discussion}

An natural question is how typical are these results?  The mean value
of the observed beam measured using Argus during commissioning
observations in good conditions (TGBT15A\_901\_34) is
$(1.15\pm0.03)\,\lambda/{\rm D}$ at 86\,GHz.  This is based on 17
independent measurements with a $1\sigma$ scatter around the mean
corresponding to $0.11\,\lambda/{\rm D}$.  Including additional
measurements from the pointing scans taken during the early Argus
science programs (2016 December -- 2017 March) over a range of
conditions, the typical beam size observed is
$(1.20\pm0.12)\,\lambda/{\rm D}$.  This result is based on 35
observations of bright quasars (3C273, 3C454.3, and 3C84) for a
frequency range of 86--90 GHz.  The range of values measured was
1.03--$1.49\,\lambda/{\rm D}$.  Therefore, the results obtained here
at 109 GHz (Figure~2) are not unusual, which is encouraging for future
high-frequency observations with the GBT.  Extrapolating the results
to the CO(1-0) rest frequency of 115.271\,GHz, one could expect to
achieve a resolution of about 6.2 arcsec in good conditions at the
upper frequency end of the Argus operational range.

\section{Concluding Remarks}

Observations taken 2017.03.23 with Argus at 109\,GHz show no
measurable degradation of the beam with respect to low frequency
observations.  The elevation scan at 109 GHz (Figure~2) shows a beam
size of $\theta_{\rm FWHM} = 6.4^{\prime\prime} (1.13\,\lambda/{\rm
  D})$.  This is the smallest beam observed to date with the GBT.
Although these observations may have benefited from better than
average conditions, the results highlight the potential of the GBT at
its highest operational frequencies, 100--115\,GHz.

I acknowledge and greatly appreciate the efforts of the staff at the
Green Bank Observatory who continue to maintain and improve the
performance of the GBT at high frequency.  I thank Jay Lockman for
providing useful comments for the memo.

\section{References}

\begin{enumerate}

\item The GBT 4mm Receiver: http://www.gb.nrao.edu/4mm/

\item MUSTANG-2: http://www.gb.nrao.edu/mustang/

\item Argus:  http://www.gb.nrao.edu/argus/

\item Condon, J.J. \& Balser D.S. 2015, Dynamic Scheduling
  Algorithms, Metrics, and Simulations, DS Project Note 5.6

\item Maddalena, R.J. \& Frayer, D.T., 2014, Recommended Changes to the
  Constants Used by the DSS for Fall 2014, DS Project Note 18.1

\item Schwab, F.R. \& Hunter, T.R. 2010, Distortions of the GBT Beam
  Pattern Due to Systematic Deformations of the Surface Panels, GBT
  Memo \# 271

\item Sieth, M. 2016, Argus: A 16-pixel Millimeter-wave Spectrometer,
  Ph.D. Thesis, Stanford University

\item Maddalena, R.J. 2010, Theoretical Ratio of Beam Efficiency to
  Aperture Efficiency, GBT Memo \#276

\item Frayer, D.T., et al. 2015, The GBT 67-93.6 GHz Spectral Line
  Survey of Orion-KL, AJ, 149, 162

\item Maddalena, R.J., Frayer, D.T., Lashley-Cohirst, N. \& Norris,
  T. 2014, The Updated 2014 Gravity Model, PTCS Project Note 76.1

\item Norrod, R. \& Srikanth, S. 1996, A Summary of the GBT Optics
  Design, GBT Memo \#155

\item ALMA Calibrator Source Catalog:
  https://almascience.eso.org/sc/

\end{enumerate}

\begin{figure}[t]
\includegraphics[width=1.0\textwidth]{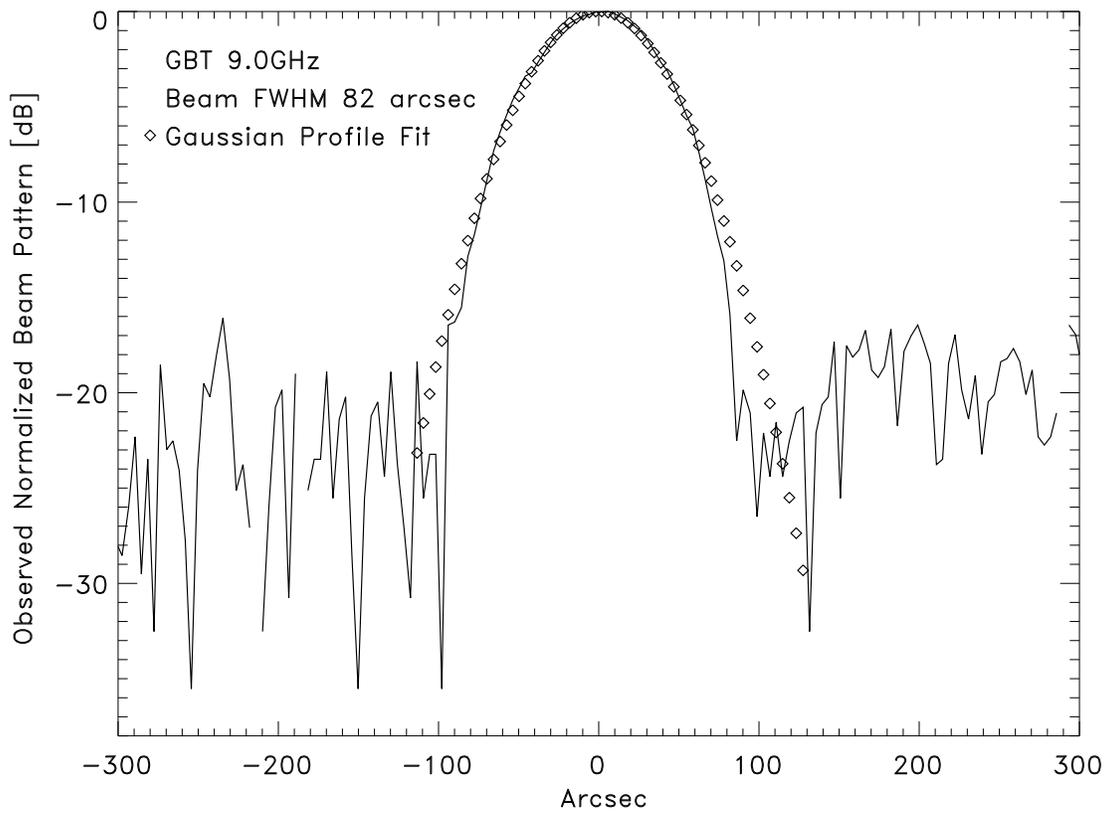}
\vspace*{-3cm}
\caption{The normalized GBT beam profile in dB observed at 9.0 GHz
  shown as the solid line as a function of arcsec from 0510+1800 along
  the elevation axis.  The diamonds show a Gaussian fit to the
  observed profile.  The absolute value of the data was used for
  plotting on a logarithmic scale to show the relative noise level of
  the observations ($-20$dB).}
\end{figure}
 
\begin{figure}[t]
\includegraphics[width=1.0\textwidth]{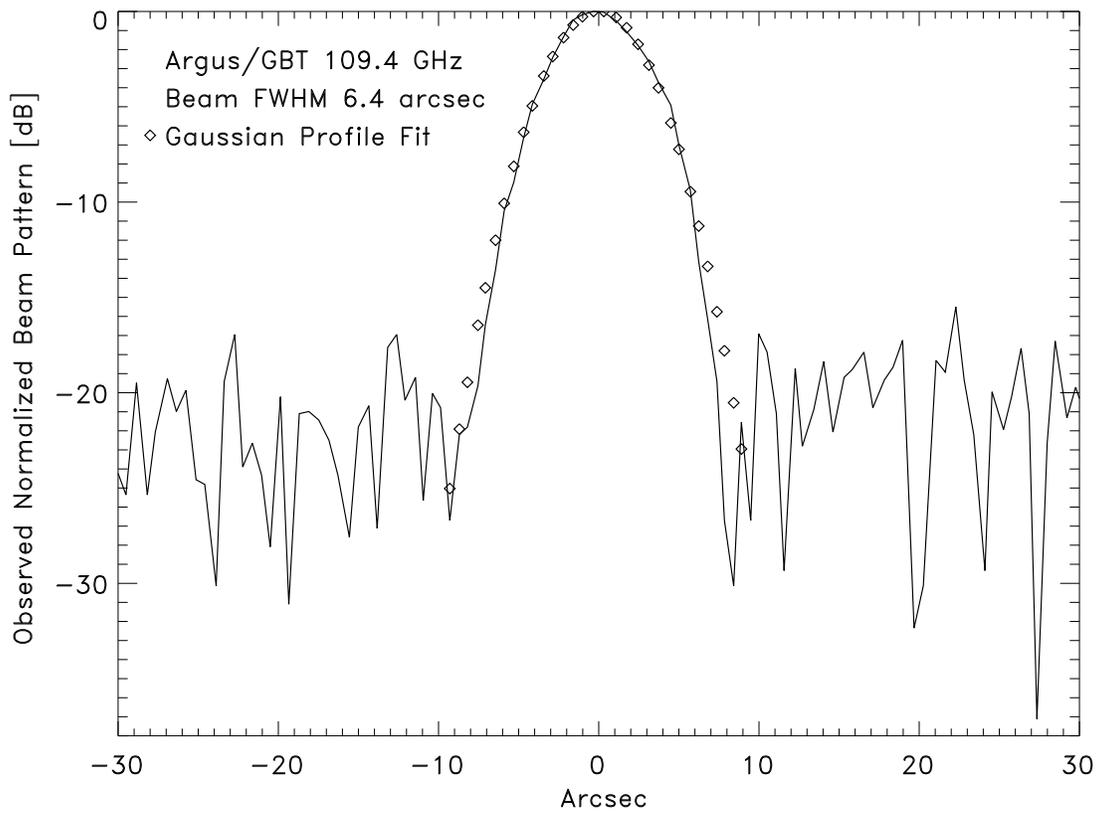}
\vspace*{-3cm}
\caption{The normalized GBT beam profile in dB observed at 109.4\,GHz
  shown as the solid line as a function of arcsec from 3C273 along the
  elevation axis.  The diamonds show a Gaussian fit to the observed
  profile.  The observed beam of 6.4 arcsec is the smallest beam
  observed to date with the GBT.}
\end{figure}

\end{document}